# A Novel Method to Obtain Wires Distribution Considering the Shape of Generated Electromagnetic Field

Tianqi Hong[1], *Student Member, IEEE*

[1]Engineering School, New York University, Brooklyn, NY 11201, USA, th1275@nyu.edu

**This paper proposes a method to calculate the wires distribution for generating required electromagnetic field. Instead of solving the distribution of wires directly, we formulate the problem into zero-one programming form. By applying the proposed algorithm to solve the zero-one programming problem, a practical solution can be obtained. Two practical examples are proposed to illustrate detailed calculation steps of the novel method. The comparison between binary particle swarm optimization searching algorithm and the proposed algorithm is provided and discussed. All the design results are validated with FEM calculation results.**

## I. INTRODUCTION

ELECTROMGNETIC field design is necessary in various applications [1]-[4]. To satisfy certain requirements, designing a particular electromagnetic field is the common objective. Numerous algorithms are developed in the past decades to reach those requirements in diverse ways.

To generate a uniform electromagnetic field above the coils for wireless power transmission, a fit function to calculate the turn distribution of a set of concentric circular coils is proposed in [1]. Instead of concentric circular coils structure, a coils array structure has been proposed in [2] to achieve uniform charging electromagnetic field for power transfer of biomedical devices. In [3], a 3-D concentric circular coils structure has been studied to generate uniform electromagnetic field for electric vehicles charging lot. To obtain a sinusoid electromagnetic field in air-gap of a superconducting synchronous generator, the particle swarm optimization (PSO) algorithm is applied in [4].

Although the requirements and applications of the electromagnetic field design in [1]-[4] are different, the objectives – design particular shapes of electromagnetic fields – are similar. In this paper, these similar electromagnetic field design problems are generalized into optimization problems. A mathematic formulation of this optimization problem is proposed and rewritten into zero-one programing form. A practical numerical method is provided to solve the zero-one programing problem. The comparison between the proposed method and PSO method is discussed. The numerical examples are given to provide the detail interpretation of the proposed method. All optimization results are validated by FEM calculation.

## II. PROBLEM DESCRIPTION

This paper is aiming to obtain power wires distribution for generating required electromagnetic field. By distributing the wires with identical current $i(t)$ on purpose, a required electromagnetic field can be obtained (say design the air-gap electromagnetic field of synchronous machine).

According to the practical applications, the electromagnetic field in particular position needs to be precisely designed to achieve better characteristics. For clear illustrate the problem, a generalized electromagnetic field design example is plotted in Fig. 1.

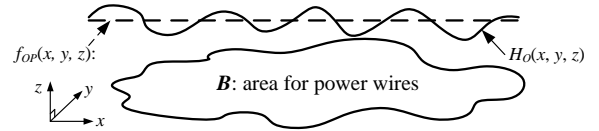

Fig. 1. A generalized example of the electromagnetic field design.

According to the example in Fig. 1, several functions are defined in 3-D coordinate $(x, y, z)$ as follows.

Objective position function $f_{OP}(x, y, z)$: A function describes the position of expecting electromagnetic field. $\boldsymbol{O}$ is the set of the discrete points which are satisfied with $f_{OP}(x, y, z)$.

Objective function of electromagnetic field $\vec{H}_O(x, y, z)$: A function represents the expecting electromagnetic field.

Placement matrix $\boldsymbol{B}$: A matrix describes the discretized area of wires placement information. The elements in Matrix $\boldsymbol{B}$ are the possible wire position vectors $\boldsymbol{p}_i = (p_x^i, p_y^i, p_z^i)$.

Optimal position matrix $\boldsymbol{P}^*$: A matrix describes the optimal distribution of powered wires to achieve the objective electromagnetic field.

Apart from the definitions above, the electromagnetic field strength function of $i^{th}$ single wire is denoted as $\vec{H}^i(x, y, z, \boldsymbol{p}_i)$. For linear problem, Biot-Savart Law can be applied to calculate $\vec{H}^i(x, y, z, \boldsymbol{p}_i)$. In nonlinear cases without considering saturation, finite element method can be applied to obtain the discrete function of $\vec{H}^i[x_k, y_k, z_k, \boldsymbol{p}_i]$.

Based on the definitions above, the electromagnetic field design problem can be equivalent into an optimization problem which is:

$$\boldsymbol{P}^* = \underset{\boldsymbol{p}_i \, for \, i=1,2,\dots,N}{\operatorname{argmin}} \left\| \vec{H}_O - \vec{H}_\Sigma \right\|_2$$
$$s.t.: \boldsymbol{p}_i \in \boldsymbol{B} \ and \ (x, y, z) \in \boldsymbol{O} \quad (1)$$
$$\vec{H}_\Sigma = \sum_{i=1}^N \vec{H}^i(x, y, z, \boldsymbol{p}_i)$$

where $\vec{H}_\Sigma$ is the summation of electromagnetic field generated by each wire and $N$ is the number of wires which is a variable in this optimization problem.

The error between the optimization result and objective electromagnetic field is defined as:

$$e = \left\| \vec{H}_O - \vec{H}_\Sigma \right\|_2 \qquad (2)$$

## III. MODEL SIMPLIFICATION AND NUMERICAL ALGORITHMS

### A. Model Simplification

Due to the nonlinearity of the problem described in Section II, the analytical global optimum is difficult to obtain. Instead of computing the $P^*$ directly, we can use an indicator matrix $A$ to present the optimal position matrix. $A$ is a "0/1" matrix which represents the existence of the wire in matrix $B$. If the wire exists in the $P^*$, the correspond element in $A$ is "1". Hence, the matrix $P^*$ can be rewritten mathematically as:

$$P^* = A_{n\times m} \cdot B_{n\times m} \qquad (3)$$

where $A_{n\times m}$ is a $n\times m$ matrix formed according to the store sequence of matrix $B_{n\times m}$.

According to (3), matrix $A$ is the only design parameter we need to be calculated. The calculation of power wire position is translated approximately into compute the distribution of "1" elements in matrix $A$. Hence, problem (1) is equivalent into a zero-one programing problem. To reduce the error of this model simplification, large $n$ and $m$ are expected.

### B. Inverse Searching Algorithm (IS)

The global optimal solution of zero-one programing problem can be obtained by exhaustive searching method. However, exhaustive searching method is not a feasible method to large scale design problem. Meanwhile, the global optimum may not a practical power wire distribution solution. A practical $A$ matrix is expected to have two properties: 1). Most of the elements in $A$ are "1"s; 2) The "1" elements are in group. Hence, the "inverse searching" (IS) algorithm is proposed to obtain a practical solution.

Different from other intelligent searching algorithm, the initial matrix $A^0$ is set to be "all-one" matrix ($a_{ij} = 1$ for all $i$, $j$). Starting from $A^0$, the correlations between each power wire and error are computed by the IS algorithm, iteratively. At each iteration step, the power wire with highest relevance is removed and $A$ matrix is updated correspondingly. The searching process will stop when the number of removed wire reaches to a value set at the beginning or the error computer from (3) stops decreasing. The flow chart of IS algorithm is represented in Fig. 2.

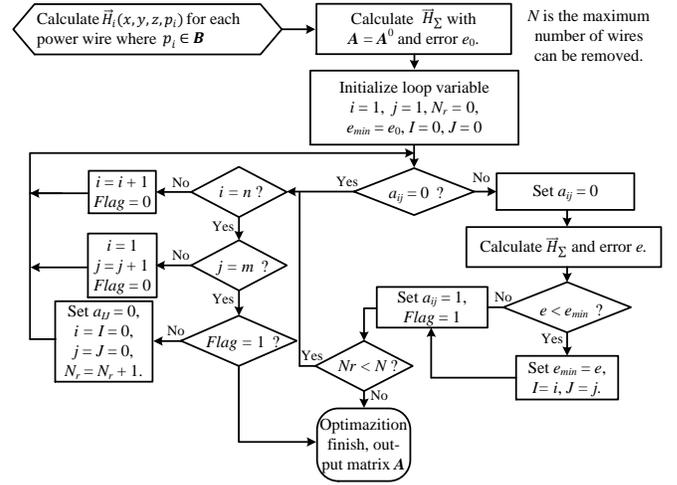

Fig. 2. Flow chart of the IS algorithm.

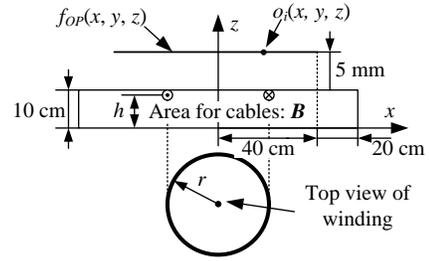

Fig. 3. The graphical model of wireless charging lot for cellphones.

### C. Binary Particle Swarm Optimization Searching Algorithm (BPSO)

The regular particle swarm optimization (PSO) algorithm is a well-studied searching algorithm which is mainly applied to solving nonlinear nonconvex problem. To solve zero-one programing problem, a modified binary PSO algorithm is chosen.

According to (3), each row vector in "0/1 matrix" $A$ can be considered as binary code, which can be translated into a decimal number. Hence, the binary matrix $A_{n\times m}^{(2)}$ is represented by a decimal vector $A_{n\times 1}^{(10)}$ and the number of variables of magnetic field design problem can be reduced from $n \times m$ to $n$. The iteration equations for regular PSO method have been modified as:

$$v_{i,d}^{k+1} = v_{i,d}^k + c_1\xi_{i,d}^k\left(Pr_{i,d}^k - x_{i,d}^k\right) + c_2\eta_{i,d}^k\left(Pg_{i,d}^k - x_{i,d}^k\right),$$
$$x_{i,d}^{k+1} = round\left[x_{i,d}^k + v_{i,d}^{k+1}\right], \qquad (4)$$

where $d = 1,\ldots,n$; $\xi^k \sim U(0,1)$ and $\eta_{i,d}^k \sim U(0,1)$ represent two random variables uniformly distributed in [0,1]; $c_1$ and $c_2$ are the acceleration coefficients; $Pr_{i,d}^k$ is the best position found by particle $i$ up to "time" $n$, and $Pg_{i,d}^k$ is the "global" best position found by particles other than $i$; $round[\cdot]$ denotes the round off function.

## IV. Application Examples

### A. Wireless Charging Lot of Cellphones

To obtain uniform charging efficiency of wireless charging lot (one transmitter with $N$ receivers), a uniform electromagnetic field along $z$-axis direction is expected.

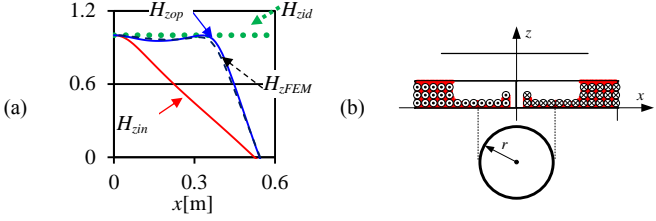

Fig. 4. Optimization results of wireless charging lot with IS algorithm; (a) The comparison between ideal objective $H_{zid}$ and optimization result $H_{zop}$ and initial $H_{zin}$, and FEM result $H_{zFEM}$, magnetic field intensities are normalized; (b) The cable distribution computed by optimization algorithm and FEM model based on optimal cable distribution.

According to the assumption that all cellphones have similar dimensions, the graphical model of magnetic field design problem is drawn in Fig. 3. The objective position function $f_{OP}(x, y, z)$ describes a plane which is 5 mm higher than the power wire (thickness of cellphones).

Based on Biot-Savart Law, the magnetic field strength along $z$ axis of a single power circular coil can be obtained as:

$$H_Z^i(x, z, p_r^i) = \frac{I}{2\pi\sqrt{(p_r^i + x)^2 + z^2}} \left[ \frac{(p_r^i)^2 - x^2 - z^2}{\left((p_r^i) - x\right)^2 + z^2} E + K \right] \quad (5)$$

where the origin is set as the center point of circular coil; $x$ is the horizontal distance between a given point in the field and the $z$ axis, $z$ is the vertical distance between a given point and the $x$ axis, $p_r^i$ is the radius of the power coil, $K$ is the elliptic integral of first kind:

$$K(k) = \int_0^{\frac{\pi}{2}} \frac{1}{\sqrt{1 - (k\sin\alpha)^2}} d\alpha \quad (6)$$

and $E$ is the elliptic integral of second kind:

$$E(k) = \int_0^{\frac{\pi}{2}} \sqrt{1 - (k\sin\alpha)^2} d\alpha \quad (7)$$

with $k = \sqrt{4p_r^i x/(p_r^i + x)^2 + z^2}$.

To consider free positioning of power coil, a position tranlation function is needed to obtain the relative position bewteen the power coil and given point. Substituting $\vec{H}^i$ in (1) by (5) with position tranlation function and applying the proposed IS algorithm, the coils distribution can be obtained; see red area in Fig. 4(b). The dimension of $A$ matrix in IS al-gorithm is 20×50, which means there are up to 1000 variables in this zero-one programing problem. The comparison between objective electromagnetic field and optimization result are plotted in Fig. 4(a).

According to Fig. 4(a), the objective electromagnetic field is achieved by IS algorithm. By distributing the coils as shown in Fig. 4(b), a uniform electromagnetic field can be obtained within a circle with radius equaling to 0.4 m.

Based on the coils distribution shown in Fig. 4(b), the FEM model is built. The FEM calculation result is plotted in Fig. 4(a) which is almost overlapped with IS algorithm result.

To compare the proposed method with intelligent searching algorithm, the uniform magnetic field design problem is also solved by BPSO method. According to BPSO algorithm proposed in Section III, it is impossible to apply BPSO algorithm because of the large dimension of $A$ matrix. The dimension of $A$ matrix is 20×50, which means there are 20 variables for BPSO algorithm and each variable goes up to $2^{50}$. Hence, the dimension of $A$ matrix is reduced to 15×25 (375 coils). By applying BPSO algorithm, the coils distribution can be obtained; see red area in Fig. 5(b) and the comparison between objective electromagnetic field and optimization result are plotted in Fig. 5(a).

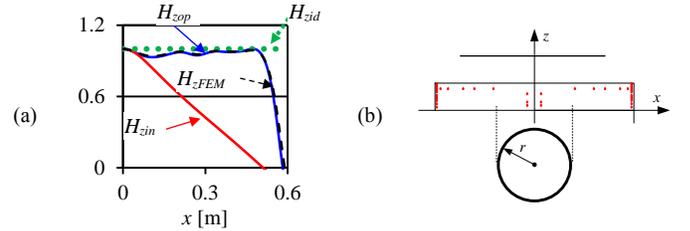

Fig. 5. Normalized optimization results of wireless charging lot by BPSO algorithm; (a) The comparison between ideal objective $H_{zid}$ and optimization result $H_{zop}$ and initial $H_{zin}$, and FEM result $H_{zFEM}$; (b) The cable distribution.

According to the comparison between Fig. 4 and Fig. 5, several conclusions can be obtained as follows.

1). The optimization result of BPSO algorithm is better than IS algorithm. The BPSO optimal result has wider uniform range (radius is 0.5 m) and small error computed from (2).

2). The BPSO algorithm is slower than IS algorithm. With increasing of the dimension of "0/1 matrix" $A$, the computation expense of BPSO algorithm increases dramatically. Hence, the proposed BPSO algorithm cannot be applied into a problem with large "0/1 matrix" $A$.

3). The absolute magnetic field strength of BPSO solution is much smaller than IS solution. Hence, the proposed BPSO algorithm is not suitable to be applied into the problem with the requirement of the amplitude of designed electromagnetic field.

4). Because of the sparse power wires distribution, the supporting structure of power wires in BPSO solution is more complex than IS solution. Hence, the manufacturing expense of BPSO solution is much higher than IS solution.

### B. Superconducting Synchronous Generator (SSG)

Due to high current density of superconductor, the superconducting synchronous generators with large capacity are intended to be designed as ironless or half ironless structure. With symmetrical ferromagnetic shielding or ferromagnetic stator, the nonlinear electromagnetic field design problems can be approximated as linear design problems. One quarter of a four-pole SSG rotor is shown in Fig. 6.

Each superconductor winding can be approximated as a rectangle coil. The rectangle coil can be decomposed into four finite long power wires. For 2-D analysis, the power wires along $x$ axis are neglected in this example. The power wires along y axis are considered as infinity length wires (neglect end effect). According to 3-D coordinate in Fig. 6, the magnetic field strength in $z$ axis of infinite long power wires along y axis is:

$$\vec{H}_z^i(x, z, p_x^i, p_z^i) = \frac{I(p_x^i - x)}{2\pi[(p_x^i - x)^2 + (p_z^i - z)^2]} \tag{8}$$

The magnetic field strength in $x$ axis of infinite long power wires along y axis is:

$$\vec{H}_x^i(x, z, p_x^i, p_z^i) = \frac{I(p_z^i - z)}{2\pi[(p_x^i - x)^2 + (p_z^i - z)^2]} \tag{9}$$

By adding the magnetic field strength of four power wires, the total magnetic field strength generated by single rectangle coil can be obtained. The magnetic field strength of rectangle coil along normal direction ($\vec{n}$) of air-gap can be computed by:

$$\vec{H}_n^i = (x^2 + z^2)^{-0.5}(z\vec{H}_z^i + x\vec{H}_x^i) \tag{10}$$

The sinusoidal magnetic field is expected along normal direction of air-gap. According to this requirement, the $\vec{H}_O(x, z)$ can be written as:

$$\vec{H}_O(x, z) = H_m \cos\{2\arccos[z(x^2 + z^2)^{-0.5}]\} \tag{11}$$

According to (10) and (11), the optimization result can be obtained by IS algorithm, where the dimension of $A$ matrix is set as $50 \times 50$. The final optimization result is shown in Fig. 7.

According to the result shown in Fig. 7, an applicable result of the power wires distribution can be obtained by proposed the novel algorithm, which generates the magnetic field in the position we expect.

To reduce the size of SSG, a high air-gap electromagnetic field is another requirement of field winding design. Hence, the proposed BPSO algorithm is not suitable for this example.

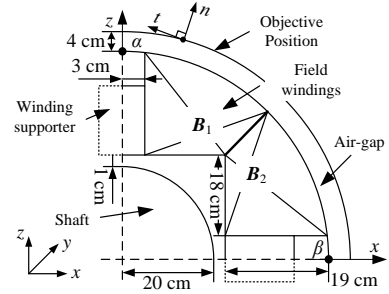

Fig. 6. One quarter of the example rotor model.

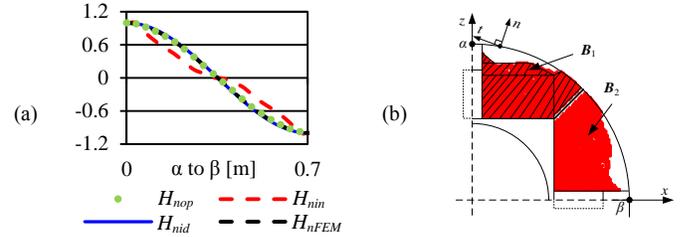

Fig. 7. Optimization result of SSG. (a) The normalized $H_n$ along arc $\alpha$ to $\beta$; blue solid line is ideal air-gap $H_{mid}$; red dash line is initial $H_{nin}$, green dots are the optimization result of $H_{nop}$ obtained from IS method, black dash line is the FEM result of $H_{nFEM}$ with magnetic shielding (overlapped with IS result); (b) The distribution of power wires, the hatched area is FEM model approximated from red area.

## V. Conclusion

In this paper, we investigated the electromagnetic field design problem. A zero-one programming formulation is proposed to describe wire distributions related electromagnetic field design problems. A novel numerical algorithm is proposed intended to solve the zero-one programming problem. By solving the design problem with IS and BPSO algorithms, the requirements of electromagnetic field can be achieved. Two practical examples are chosen to illustrate the proposed algorithm and the optimization results of both algorithms are validated by FEM calculations.